\documentclass[preprint,,showpacs,preprintnumbers,amsmath,amssymb]{revtex4-2}

\usepackage{graphicx}
\usepackage{epstopdf}
\usepackage{dcolumn}
\usepackage{bm}
\usepackage{amsmath}
\usepackage{graphicx} 
\usepackage{dcolumn}
\usepackage{bm}
\usepackage{amsmath}
\usepackage{epsfig}
\usepackage{float}
\usepackage{epstopdf}
\begin{document}

\pagebreak
	
	\title{
		Terahertz-nanoscale visualization of the microscopic spin-charge architecture of colossal magnetoresistive switching
		}
	
	\author
	{Samuel Haeuser $^{1,2}$, Randall K. Chan$^{1,2}$, Richard H. J. Kim$^{2}$, Joong-Mok Park$^{2}$, 
    Martin Mootz$^{2}$,
    Thomas Koschny$^{2}$, 
    Jigang Wang$^{1,2\dag}$
  		}
	\affiliation{$^1$ Department of Physics and Astronomy, Iowa State University, Ames, Iowa 50011, USA. \\$^2$Ames National Laboratory of U.S. Department of Energy, Ames, Iowa 50011, USA. \\
\noindent$^{\dagger}$Corresponding author: jgwang@ameslab.gov }   
	
	\date{\today}
	
	\begin{abstract}
Resolving sub-10 nm spin switching and the associated terahertz (THz) electrodynamics during the colossal magnetoresistance (CMR) transition is a definitive frontier in reaching the fundamental spatial, temporal, and energy-dissipation limits of spin-based microelectronics and quantum logic architectures. 
Yet, the requirement of simultaneous control of high magnetic field, cryogenic environment, and nanometer-sacale resolution has remained an elusive benchmark for terahertz nanoscopy, leaving the obscured nano-scale high-frequency dynamics of these transitions largely unexplored.
Here, we overcome these limitations by utilizing a custom-built cryogenic magneto-THz scattering-type scanning near-field optical microscopy (cm-THz-sSNOM) platform to resolve the nanoscale, THz spectroscopic evolution of the magnetic field-driven CMR transition in a manganite single crystal $\text{Pr}_{2/3}\text{Ca}_{1/3}\text{MnO}_{3}$.  
Our measurements provide a real-space visualization of the local THz conductivity, capturing the moment that magnetic-field-induced spin switching triggers the phase transition from an antiferromagnetic insulator to a ferromagnetic metal.
THz nano-imaging, together with an ellipsoidal near-field model, reveals a multi-scale transition initiated by 1-2 nm isolated spin-flip sites at low magnetic fields, which coalesce into $\sim$15~nm conducting regions as the threshold field is approached. 
These results provide an in situ, previously inaccessible THz real-space view of CMR switching, establishing a general analysis framework for mapping spin–charge–lattice–orbit–coupled dynamics at spatial scales that transcend the nominal sSNOM resolution.
	\end{abstract}
	
	\maketitle

	\section{Introduction}
There has been a significant recent push to establish cryogenic temperatures and high magnetic fields within scattering-type scanning near-field optical microscopy (sSNOM) \cite{kim3, jutopo, dapo, dapolito, wehmeierlandau, kim5, fei, ni, haeuser2026, kim2026} to probe correlated phenomena at nanoscale spatial resolutions. Terahertz (THz) frequencies are uniquely suited for their investigation \cite{huang2026, yang1,yang2,yang3,mootz2022} since many fundamental spin–charge–lattice–orbit–coupled excitations responsible for these phenomena reside in the low-energy ($\sim$meV) regime. However, magneto-THz spectroscopy has been predominantly restricted to the diffraction limit, averaging over complex microscopic features and leaving the underlying dynamical principles of these phase transitions incomplete. As illustrated in Fig. 1a, THz-sSNOM has already been applied to resolve local heterogeneities in a diverse array of materials--including nano-junctions, topological materials, and perovskite solar cells--demonstrating its unique capacity to uncover microscopic physics that remains hidden to far-field probes \cite{kim, kim2, kim4, kim5,kim}.
While recent instrumentation advances have established our cryogenic magneto-THz scattering-type scanning near-field optical microscopy (cm-THz-sSNOM) platform as a unique tool capable of reaching 1.8~K and multi-Tesla fields \cite{kim3, kim2026}, these capabilities have yet to be leveraged to resolve strongly correlated electronic transitions or to provide a quantitative framework for extracting microscopic physical insights from complex, real-space spin-charge coupled landscapes.

Colossal magnetoresistance (CMR) in perovskite-type manganites, $\text{R}_{1-x}\text{A}_{x}\text{MnO}_3$ (R and A are trivalent rare-earth and divalent alkaline-earth ions, respectively), represents a prominent model system for exploring this correlated physics and remains a primary driver for next-generation spintronics \cite{tomioka2001,tomioka1996PCMO,tokunaga,liAnisotrop}. The prevailing consensus for the CMR mechanism is the phase-separation model, which posits that the colossal resistance change originates from a delicate competition between {\em nanoscopic, heterogeneous} domains of antiferromagnetic (AFM) insulating and ferromagnetic (FM) metallic phases \cite{dagottoBook}. Despite this understanding, two outstanding open questions remain. First, despite the demand for probing high-frequency electrodynamics of CMR transition far exceeding the DC limit, existing THz studies remain diffraction-limited and cannot resolve the local switching dynamics. Second, while microscopy on strained thin films \cite{laiMIM,gonzalezNANO,zhangMFM} and defect-engineered crystals \cite{fathSinglecrystal} has visualized domain formation on the 50-100 nm scale, these observations are often dominated by  the external confinement rather than the intrinsic microscopic spin-charge architecture of the transition. Prior investigations in high-quality single crystals have been limited to STM and X-ray diffraction under DC and field-cooling conditions \cite{rosslerSTM,beckerSTM,dagottoOPEN,TaoNANO}. Critically, current nanoscopic techniques such as THz-STM lack the essential in-situ magnetic field tuning required to drive and resolve the real-space evolution of the magnetic swithcing.

We chose the $\text{Pr}_{2/3}\text{Ca}_{1/3}\text{MnO}_{3}$ (PCMO) as a model system to demonstrate the capability of cm-THz-sSNOM in resolving the local spin-charge heterogeneity during the CMR transition. As illustrated in Figs.~1b-1c, PCMO has $\text{MnO}_6$ orthorhombic ordered perovskite structure and hosts a complex CE-type AFM ground state characterized by zig-zag chains of alternating $\text{Mn}^{4+}$ and $\text{Mn}^{3+}$ ions as a charge-ordered and orbital-ordered (CO/OO) architecture \cite{jirakneutronNEUTRON,tomioka2001,tomioka1996PCMO,tokunaga,dagottoBook}. 
In this CO/OO state, the magnetic transition serves as the primary trigger for conductivity switching; inter-chain electron hopping is strictly constrained by Hund’s rule, which imposes a large energy penalty against the local AFM order. As illustrated in Fig.~1d, the application of an external magnetic field flips these spins into a FM state, ``unlocking" conduction pathways via the double-exchange mechanism. However, the ``colossal" magnitude of the resistance drop is fundamentally driven by the subsequent ``melting" of the CO/OO state. Jahn-Teller lattice distortions in the insulating state trap electrons within a rigid, real-space CO/OO lattice, which only collapses when field-induced spin switching acts as the catalyst for a structural phase transition. This direct mapping between magnetic order (AFM vs. FM) and local conductivity (insulating vs. metallic) provides an ideal platform for cm-THz-sSNOM to study this spin–charge–lattice–orbit–coupled excitations. While previous ensemble studies of spin dynamics have utilized this material to demonstrate ``quantum femtosecond magnetism" \cite{li2013, lingos2017}--including 100-fs photo-induced AFM-to-FM switching--those far-field ultrafast spin dynamics measurements averaged over complex microscopic features. By utilizing cm-THz-sSNOM, we move beyond these ensemble averages to probe the real-space microscopic and THz dynamic architecture of the CMR transition, visualizing the nanometer, THz spectroscopic evolution of conducting domains as the spin-charge-coupled heterogeneous domains at its fundamental spatial limits.

		\begin{figure} [hbt!]
			\centering
			\includegraphics[width=1\linewidth]{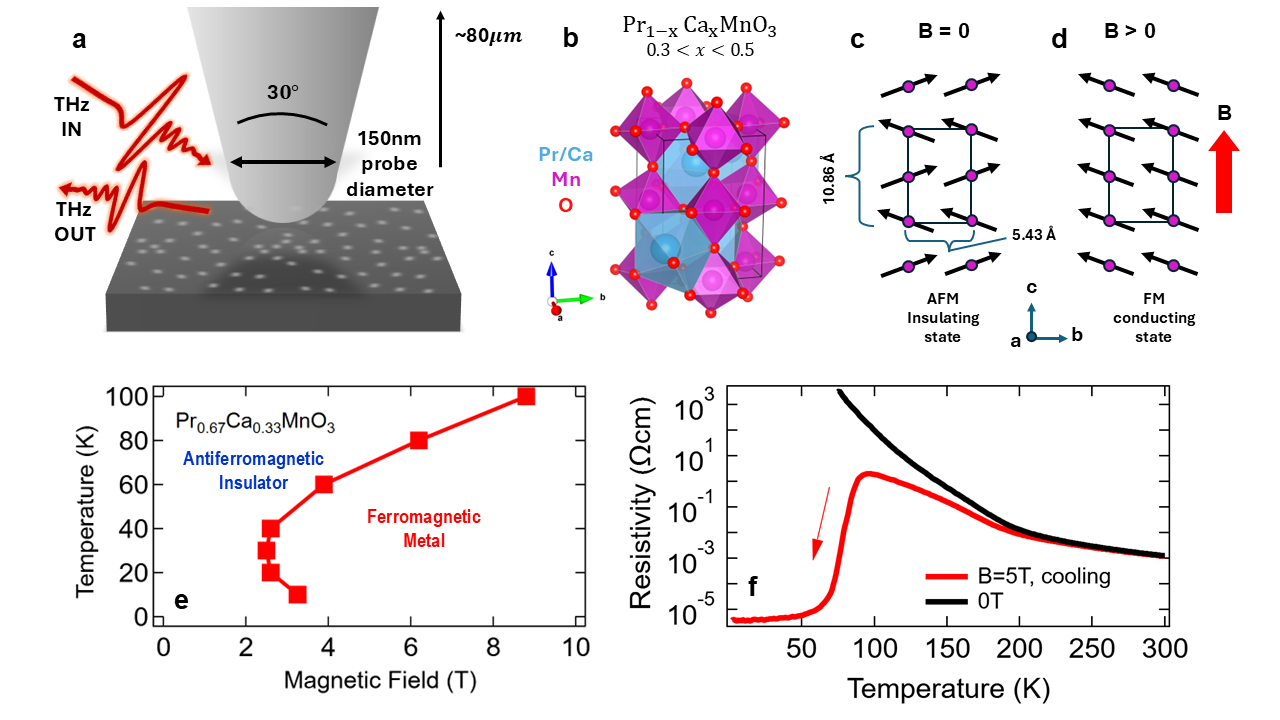}
			\caption{\textbf{Terahertz magneto-nanoscopy of the colossal magnetoresistance (CMR) phase transition.} 
\textbf{a)}~Schematic representation of the $\sim$80~$\mu$m tall, $\sim$150~nm diameter Pt-coated AFM probe above the Pr$_{0.67}$Ca$_{0.33}$MnO$_3$ surface, illustrating the detection of sub-resolution conducting and insulating domains. 
\textbf{b)}~Orthorhombic crystal lattice structure of Pr$_{0.67}$Ca$_{0.33}$MnO$_3$ \cite{Vesta3}. 
\textbf{c,d)}~Schematic of the Mn$^{3+}$/Mn$^{4+}$ spin configurations in \textbf{c)} the zero-field charge-ordered insulating state and \textbf{d)} the magnetic-field-induced charge-disordered metallic state. 
\textbf{e)}~Macroscopic magneto-transport phase diagram showing the insulator-to-metal transition boundary as a function of magnetic field and temperature. 
\textbf{f)}~Resistivity as a function of temperature for constant magnetic fields of 0~T (black) and 5~T (red), highlighting the CMR effect upon field-induced melting of the charge-ordered state. }
			\label{fig:figure1}
		\end{figure}        

In this Article, we resolve the underlying dynamical switching principles governing the colossal magnetoresistive transition by providing the THz nano-imaging of the magnetic field-induced ``melting" of the AFM insulating state in $\text{Pr}_{2/3}\text{Ca}_{1/3}\text{MnO}_{3}$. 
This material platform is chosen because it manifests a direct mapping between magnetic switching and local THz conductivity. 
Utilizing our custom-developed cm-THz-sSNOM, we monitor the evolution of the real-space mixed-phase landscape in situ as an external magnetic field drives the spin switching that trigger the CMR transition. The THz near-field nano-spectroscopy, integrated with the modeling, yields quantitative insights into the microscopic electronic structure on both sides of the transition boundary, uncovering domain THz dynamics that are inaccessibe by previous DC and low-frequency microscopy probes. Significantly, this framework can be extended to a wide range of strongly correlated electronic transitions and spin-charge quantum excitations that require simultaneous nanometer-scale spatial resolution and THz frequency sensitivity.

\section{Results}

The charge-ordered/orbital-ordered (CO/OO) state robustly forms in $\text{Pr}_{1-x}\text{Ca}_{x}\text{MnO}_3$ for the doping level $0.3 < x < 0.5$ \cite{tomioka1996PCMO}. Figs.~1e-1f show macroscopic magneto-transport characterization of our $\text{Pr}_{2/3}\text{Ca}_{1/3}\text{MnO}_{3}$ single crystal to establish a baseline for its CMR switching behavior. In the absence of an external magnetic field, the crystal remains in an insulating state across the entire temperature range (black solid line, Fig.~1f), as the CO/OO electronic structure stabilizes the AFM spin order. The application of a magnetic field facilitates the FM reordering of these spins, which in turn acts as the catalyst to disorder the CO/OO lattice and trigger a structural phase transition into the metallic state (red solid line, Fig.~1f). Temperature-dependent resistivity measurements reveal the formation of the AFM insulating state at low temperatures, while field-cooled cycles demonstrate a second-order transition to a highly metallic phase as magnetic fields increase (Fig.~1e). 
Magnetoresistance curves at a fixed temperature after zero-field cooling show that the system's resistance drops by several orders of magnitude upon reaching a critical field (Fig.~1f). According to the measured phase diagram for this doping concentration in Fig.~1e, the ideal temperature for inducing a transition at the lowest possible field between 2-3~Tesla after zero-field cooling occurs at $\sim$30 K \cite{tomioka1996PCMO}. This specific thermodynamic window was selected for our in-situ cm-THz-sSNOM imaging to maximize the sensitivity to field-induced domain dynamics.

		
		\begin{figure} [hbt!]
			\centering
			\includegraphics[width=1\linewidth]{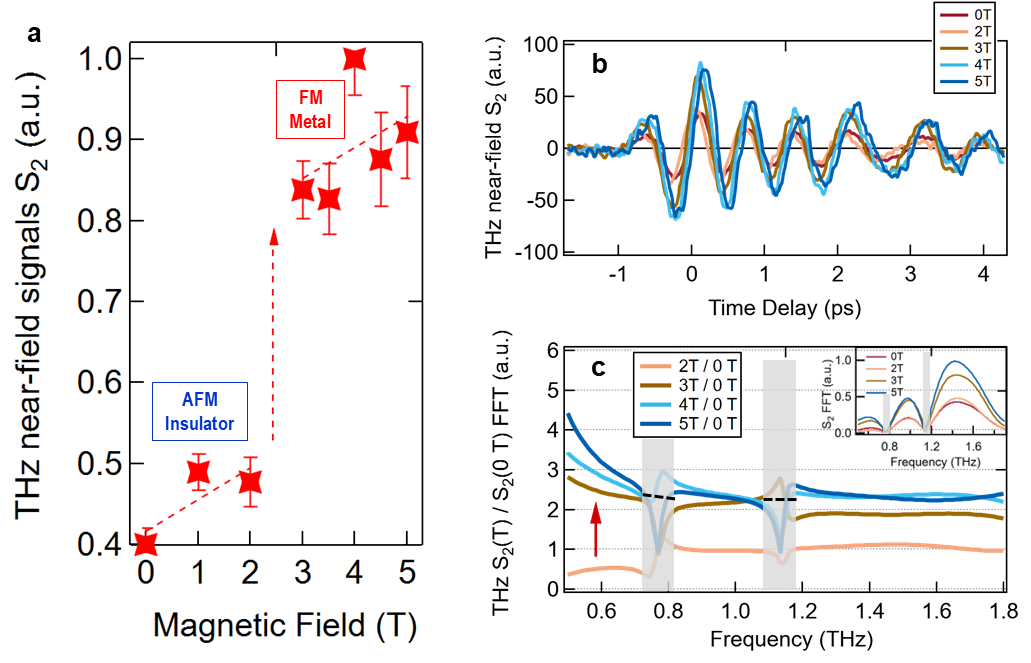}
			\caption{\textbf{Magnetic field-driven THz near-field nano-spectroscopy and spectral evolution.} 
\textbf{a)}~Integrated THz near-field $S_2$ peak amplitude as a function of increasing magnetic field following zero-field cooling to 29~K. The dashed line illustrates the insulator-to-metal transition (IMT) triggered by AFM-FM spin-switching. 
\textbf{b)}~Representative time-domain THz $S_2$ near-field waveforms recorded at discrete magnetic fields (0~T, 2~T, 3~T, 4~T, and 5~T), capturing the field-induced enhancement of the scattered electric field. 
\textbf{c)}~Normalized THz $S_2$ spectral ratios ($S_2(B)/S_2(0~T)$) revealing the emergence of a Drude-like response and low-frequency enhancement above the threshold field. Grayed regions denote instrumental artifacts arising from tip-cantilever resonances. 
\textbf{(Inset)}~Corresponding raw THz $S_2$ spectra derived via Fourier transform of the time-domain traces in \textbf{(b)}.}
			\label{fig:figure1}
		\end{figure}

To investigate the real-space THz electrodynamics of the CMR transition, we employed a custom-developed cm-THz-sSNOM platform operating at 29~K in magnetic fields up to 5~T (Method). By focusing THz fields onto a metal-coated AFM tip, we utilize the probe as a nanoscopic antenna to confine incident radiation and amplify near-field scattering at the tip-apex. With the tip positioned at a fixed coordinate on the PCMO surface, we perform time-domain THz nanospectroscopy by scanning a motorized delay stage; this modulates the arrival of the optical sampling pulse at the electro-optic crystal to resolve the oscillating electric-field waveform of the scattered near-field amplitude.The near-field signals ($S_n$) are extracted by demodulating the backscattered radiation at the $n$-th harmonic of the tip-tapping frequency ($n=2$) to effectively suppress far-field background. We performed zero-field cooling to 29 K to initialize the system in the highly ordered CO/OO AFM insulating state, subsequently ramping the magnetic field while performing in-situ time-domain THz nanospectroscopy on the PCMO surface. Any fluctuations in tapping amplitude due to the magnetic field were accounted for via an in-situ reference protocol (Supplementary). 

By fixing the sampling delay at the peak of the THz waveform (Fig.~2b), we obtained the THz near-field peak amplitude ($S_2$) as a function of the external magnetic field up to 5~T (Fig.~2a). Figure~2a displays the peak THz near-field signal ($S_2$) as a function of the increasing magnetic field. We observe a distinct increase in the THz scattered signal between 2~T and 3~T, marking the insulator-to-metal transition (IMT) as the magnetic field initiates spin flipping and triggers the collapse of the AFM-CO/OO state. This ``sharp transitional region” corresponds to the field range in which the majority of spins undergo collective switching and reorganize into the FM metallic state. This behavior is directly evidenced, for the first time, by the near-field THz time-domain traces shown in Fig.~2b, which capture the transition beyond the capabilities of conventional DC magneto-transport and other low-frequency techniques employed to date.


Furthermore, we performed a detailed local THz electrodynamics measurement of the CMR phase transition by examining the near-field THz spectral response, obtained via Fourier Transform of the time-domain waveforms at varying magnetic fields (Fig. 2b). 
A Fourier transform of the time-domain waveforms yields the local spectroscopic response, spanning a broad frequency range from 0.5 to 1.8~THz as shown in Fig.~2c. To isolate the electronic evolution, the spectra at 2~T, 3~T, 4~T, and 5~T are normalized to the 0~T insulating state (Fig.~2c). Two spectral regions (centered at $\sim$0.8 THz and 1.2 THz) have been grayed out to avoid misinterpretation; these large fluctuation feature are artifacts arising from low spectral weight in the 0~T reference, caused by intrinsic tip-cantilever resonances (Fig. 2c, inset). The resulting relative spectra reveal a systematic evolution of the local THz electrodynamics across the CMR transition. Below the threshold field, the 2~T spectral ratio (orange solid line) is characterized by a flat, featureless response, which is indicative of a frequency-independent dielectric response within the charge-ordered insulating phase. As the magnetic field exceeds the transition boundary, a distinct Drude-like spectral signature emerges, marked by a significant low-frequency enhancement-or ``overshoot"-below $\sim$0.5 THz. This low-frequency rise increases with the magnetic filed (i.e., 3~T vs 4~T vs 5~T traces) which signifies the emergence of mobile charge carriers and the increased occupation of a metallic state. In the high-frequency regime (above $\sim$1~THz), the spectra remain relatively flat but exhibit a consistent field-dependent increase in magnitude. This overall spectral behavior can be understood as a real-space convolution of emerging metallic domains within the insulating matrix, where the THz near-field probe captures the integrated response of the heterogeneous phase-separated landscape.

	
		\begin{figure} [hbt!]
			\centering
			\includegraphics[width=1\linewidth]{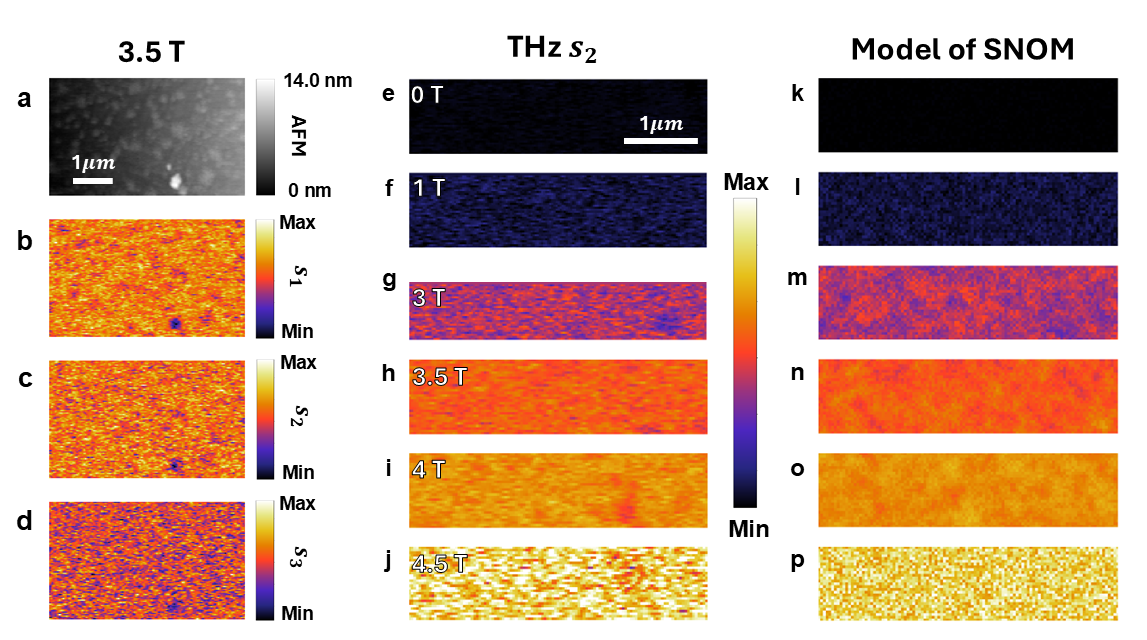}
			\caption{\textbf{Real-space THz nano-imaging and sub-resolution modeling of the CMR transition.} 
\textbf{a--d)}~Simultaneous acquisition of \textbf{a)} AFM topography, \textbf{b)} THz $s_1$, \textbf{c)} THz $s_2$, and \textbf{d)} THz $s_3$ near-field signals over a $5 \times 4$~$\mu$m$^2$ region at 3.5~T. The topographic mapping confirms a flat surface profile, ensuring that the observed near-field contrast is electronic in origin. 
\textbf{e--j)}~Experimental second-harmonic ($s_2$) THz near-field images as a function of magnetic field (0~T to 4.5~T) on a $4 \times 1$~$\mu$m$^2$ global scale. 
\textbf{k--p)}~Corresponding modeled THz near-field images generated via an ellipsoidal dipole model and stochastic spin-flip population, reproducing the experimental signal evolution and spatial homogeneity across the transition.}
			\label{fig:figure1}
		\end{figure}

To visualize the spatial evolution of the field-driven CMR transition, we performed THz near-field imaging across a representative area of the PCMO surface. Figures.~3a-3d validate the measurement technique at 3.5~T--a field strength where the majority of the sample has transitioned into the metallic state. The atomic force microscopy topography (Fig.~3a) confirms a highly polished surface with minimal roughness ($<$4 nm peak-to-valley variation), effectively ruling out topographic artifacts in the near-field response, with the exception of an isolated $\sim$14 nm dust particle utilized for spatial registration. Simultaneous demodulation of the 1st ($S_1$), 2nd ($S_2$), and 3rd ($S_3$) harmonics (Figs.~3b–3d) demonstrates that the $S_2$ signal provides the optimal balance between high signal-to-noise ratio and effective suppression of far-field background. Consequently, $S_2$ was selected for detailed nano-imaging and subsequent modeling. 
Figures~3e–3j display the evolution of the THz $S_2$ near-field signal across a global scale as the external magnetic field is increased from 0~T to 4.5~T. Consistent with the spectroscopic point-measurements (Fig. 2a), we observe a pronounced, monotonic increase in the integrated near-field amplitude spanning the transition from the insulating state (low fields, e.g., 0~T and 1~T) to the metallic state (high fields, e.g., $>$3~T). Crucially, however, the spatial distribution within individual near-field images remains strikingly homogeneous, appearing dominated by random fluctuations rather than distinct macroscopic domain formation. Instead of resolvable segregated regions, we observe a continuum of signal rise driven by the external field. 

This spatial homogeneity suggests that the CMR transition in high-quality single crystals does not proceed via the growth of large-scale metallic islands. Instead, it supports a model where the transition is composed of a dense distribution of sites where switching has or has not occurred, leading to a macroscopic slope in the transition profile rather than a discrete jump \cite{tokunaga}. 
Since the nominal spatial resolution of our $S_2$ signal is $\sim$50 nm-as determined by established approach curves \cite{kim3} and the interaction volume analysis below—the lack of resolvable heterogeneity implies that the intrinsic architecture of the transition consists of phase-separated regions significantly smaller than this benchmark. Previous studies have reported that the phase transition can occur locally on length scales comparable to the crystal lattice constant \cite{tomioka2001,dagottoOPEN,jirakneutronNEUTRON,gonzalezNANO,rosslerSTM,beckerSTM}.
Therefore, our observations necessitate the use of sub-resolution modeling to quantify the spatial evolution of these "buried" domains at scales that transcend the nominal THz-sSNOM resolution.

\section{Discussion}

The scanning probe employed for THz nano-imaging features an estimated apex diameter of approximately $\sim$150 nm (Fig. 1a). As established in prior studies of THz near-field probes, the achievable spatial resolution does not scale linearly with the geometric tip size \cite{maiss}. Instead, when higher-harmonic demodulation is utilized, the effective spatial resolution is determined by the nonlinear near-field interaction volume, governed by the ratio $A/R$, where $A$ denotes the tapping amplitude ($\sim$150 nm) and $R$ the tip radius ($\sim$75 nm). Under these operating conditions and with $S_2$ demodulation, the effective near-field confinement allows the cm-THz-sSNOM system to resolve heterogeneous features on the order of $\sim$50 nm. The fact that our images remain spatially ``featureless" at this scale is a significant physical result: it indicates that the metallic ``seeds" of the CMR transition are significantly smaller than the tip-sample interaction volume. To access electronic textures below the nominal nanoscopy resolution limit, we employ sub-resolution modeling that explicitly accounts for tip–sample convolution and quantitatively benchmark the simulations (Figs.~3k-3p) against the experimental images (Figs.~3e-3j). This analysis extracts the underlying domain scales, revealing a transition initiated by isolated 1–2 nm switching sites that eventually coalesce into $\sim$15 nm regions elaborated below.

		\begin{figure} [hbt!]
			\centering
			\includegraphics[width=0.8\linewidth]{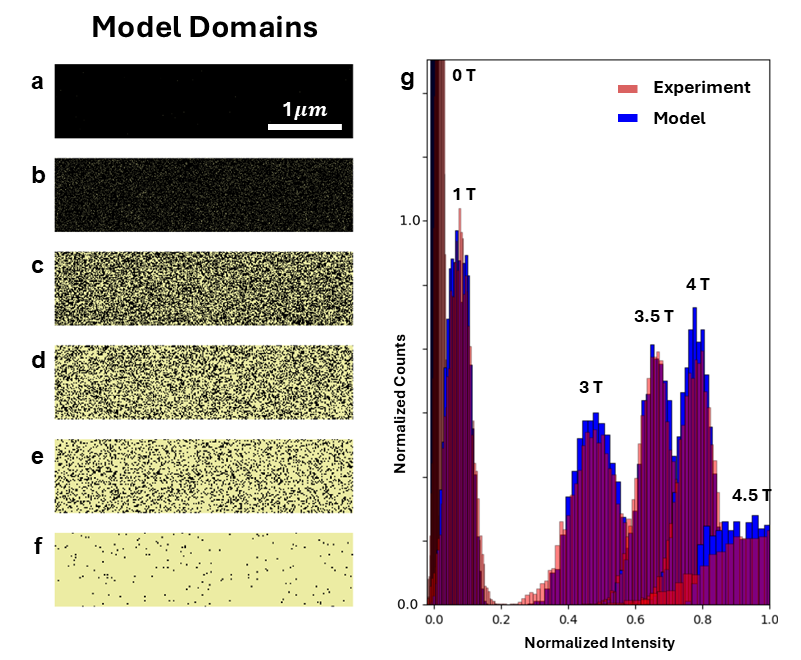}
			\caption{\textbf{Multi-scale domain evolution and statistical validation.} 
\textbf{a--f)}~Unfiltered modeled domain structures illustrating the stochastic population of conducting spin-flip sites across the magnetic field-driven transition (0~T to 4.5~T). This simulation reveals the transition's progression from isolated 1--2~nm switching sites at low fields to the formation of $\sim$15~nm percolative regions as the metallic volume fraction increases. 
\textbf{g)}~Intensity histograms comparing the spatial signal distribution of the experimental near-field images (red) and the modeled results (blue). The excellent agreement across all magnetic fields validates the use of the ellipsoidal near-field interaction model to quantify the underlying phase-separated architecture.}
			\label{fig:figure1}
		\end{figure}
	
To quantify the microscopic nature of the CMR transition, we employ a sub-resolution modeling framework that simulates the field-driven population of metallic regions within the insulating host. Following the current consensus of the multi-component spin-charge-lattice-orbit-coupled nature of this real-space transition, we consider the primary driver of the CMR conductivity change to be the individual spin-reordering event that acts as the catalyst for the subsequent structural transition and the melting of the CO/OO state. In the PCMO crystal, the fundamental length scale for this spin-flip process is the lattice constant along the $c$-axis \cite{tomioka2001,jirakneutronNEUTRON}, which places the minimum physical region for the initiation of the phase transition at approximately $\sim$0.5 nm. By populating a simulated grid with either conducting or insulating domains of varying sizes, we modeled the near-field response while incorporating experimentally measured constraints, including instrumental noise and the in situ reference protocol. To accurately account for the spatial filtering induced by a probe of a given radius, we utilized an ellipsoidal near-field model (see Supplementary Information for details) to define the largest tip-sample interaction volume. For a probe with a smooth apex, the axial distribution of the near-field interaction is found to be Lorentzian-like to the first order harmonic. By applying a blur convolution kernel derived from this Lorentzian interaction profile, we mapped the relationship between the measured $S_2$ signal and the underlying domain size distribution. This interaction model represents the first order harmonic interaction and we use this to understand the widest range of the interaction volume to provide an upper bound of the resolution. This model allowed us to extract the average estimated domain size as a function of the metallic volume fraction, based on the assumption that the density of flipped spins ``grows" spatially with the increasing magnetic field. 

To validate our sub-resolution hypothesis, we generated modeled near-field images for magnetic fields ranging from 0~T to 4.5~T (Figs. 3k–3p) to be compared directly with the experimental global-scale images (Figs. 3e–3j). In this framework, the experimental data was utilized solely to establish the average conducting area population and the background noise floor, ensuring a rigorous comparison. The resulting modeled images are strikingly consistent with the experimental data, reproducing not only the integrated signal levels but also some of the specific spatial distributions observed across the transition. Notably, the model successfully captures the emergence of subtle, stochastic ``structure-like" features--regions where local population density fluctuations survive the tip-sample convolution--closely mirroring the real-space heterogeneity observed in the experimental images. 

To resolve the underlying physical mechanism of this transition, we examined the unfiltered modeled domain structures prior to tip-convolution (Figs.~4a-4f). This scheme reveals a multi-scale evolution: at low magnetic fields (e.g., 1 T), the transition is initiated by isolated 1–2 nm switching sites, most likely representing individual spin-flip events within the AFM matrix. As the magnetic field increases, there is a heightened probability of neighboring spin realignment, which drives the ``growth" and coalescence of these conducting domains into $\sim$15 nm percolative regions. We estimate the FM metallic volume fractions to be 0.0\% (0~T), 9\% (1~T), 48\% (3~T), 66\% (3.5~T), 78\% (4~T), and 98.5\% (4.5~T), respectively. These findings indicate that the gradual signal increases observed outside the primary transition region (0-2 T) originate from independent, local spin induced phase-switching events, whereas the colossal jump near 3~T is driven by a rapid percolative expansion of these nanoscopic FM metallic domains. The robustness of this model is confirmed by the histogram analysis in Fig.~4g, which shows an excellent quantitative agreement between the experimental (red) and modeled (blue) intensity distributions to underpin the field-dependent evolution of the conducting volume fraction as it approaches the percolation threshold. This robust alignment provides a previously inaccessible window into electronic switching processes that originate at the individual lattice scale, demonstrating that the CMR transition is fundamentally defined by the evolution of sub-15 nm conducting textures. 

To the best of our knowledge, there have been remarkably few investigations into the real-space structure of the CMR phase transition in high-quality single crystals. While a local lattice-switching model has been widely assumed in earlier theoretical and ensemble studies \cite{tomioka2001,dagottoOPEN,jirakneutronNEUTRON,gonzalezNANO,tokunaga}, experimental verification has remained elusive. Prior efforts utilizing STM and X-ray diffraction \cite{rosslerSTM,TaoNANO,beckerSTM} have been largely restricted to the DC limit or static field-cooling conditions. Our results provide the first nanoscopic, in-situ THz visualization of this transition as it evolves under a continuously tuned magnetic field, finally confirming the long-standing theoretical assumption that the transition is driven by the nanoscopic reordering of individual spin-charge sites. Furthermore, the local THz electrodynamics of the CMR transition have remained unexplored at the 50~nm spatial scale until our measurement; as shown in Figures~2b--2c, our measurements reveal a distinct low-frequency spectral enhancement and magnetic-field-driven ``overshoot'' below 0.5~THz. This spectral signature provides the first real-space, THz frequency spectroscopic evidence of emerging metallic domains within the insulating matrix.

\section{Summary}

In conclusion, we have utilized cm-THz-sSNOM to resolve the real-space, THz nano-spectroscopic evolution of the CMR transition in single-crystal $\text{Pr}_{2/3}\text{Ca}_{1/3}\text{MnO}_{3}$. Our results demonstrate that the field-driven melting of the CO/OO AFM insulating state is fundamentally a multi-scale process, initiated by 1-2 nm isolated spin-flip sites that coalesce into $\sim$15 nm percolative metallic regions. By inducing this transition in situ through magnetic field tuning-rather than traditional field-cooling-we provide a definitive real-space verification of the nanoscopic switching models previously assumed for high-quality single crystals. 
This work establishes a transformative foundation for utilizing THz nanoscopy to explore correlated electronic phenomena across the full temperature-magnetic field phase space. The ability to visualize these transitions at the single-digit nanometer scales and millielectronvolt energy regime paves the way for the design of next-generation spintronics and quantum logic architectures operating at their ultimate spatial and energy-dissipation limits.

\section{Methods}
	
		The broadband cm-THz-sSNOM instrument is based on a tapping-mode atomic force microscope inside a top-loading 5-Tesla split-pair magnet cryostat that can apply fields perpendicular to the sample stage with a base temperature of 1.8~K through a helium exchange gas system. The atomic force microscope is a fiber-based cantilever system designed for applications at low temperatures and in high magnetic fields. Development and operational details of our device are described in previous works \cite{kim3, kim, kim2, kim5, haeuser, kim2026,haeuser2026}. In operation, we focus THz electromagnetic fields to a sharp metallic probe that acts as an antenna to receive and amplify the near-field interaction through field enhancement of the tip resonances. This interaction scatters near-field encoded signals to the far field for detection \cite{kim3, kim, kim2, kim4, kim5, fei, hill, htchen, ribb, mast, von, dapo, dapolito, jutopo}. Figure 1~a shows a representation of this interaction.
	
	The cm-THz-sSNOM system operates at terahertz frequencies generated via optical rectification by pumping a 2.5 mm thick GaP crystal with 90 fs, 1030 nm laser pulses from a Light Conversion Pharos-UP system. To minimize ambient water vapor absorption, the entire THz beam path is continuously purged with dry air. The near-field probe—which serves as a nanoscopic antenna to capture and scatter the electromagnetic field—is a metallic-coated cantilever (Rocky Mountain Nanotechnology, model 25Pt300B) featuring an 80 $\mu$m long Pt shaft, a nominal tip apex diameter of $\sim$150 nm, and a fundamental tapping frequency of $\sim$15 kHz.To accurately quantify the probe geometry for the ellipsoidal near-field model, the effective tip radius was determined through a rigorous analysis of the AFM mechanical response, specifically utilizing tip-sample approach curves to extract the precise interaction volume. During scanning, the scattered near-field radiation is collected and resolved via electro-optic (EO) sampling. The $s_n$ signals are extracted by demodulating the backscattered field at the $n$-th harmonic of the tip-tapping frequency ($n=2$); these signals are simultaneously processed and averaged using a lock-in amplifier at each pixel to ensure high-fidelity spatial mapping.
	
With the probe in contact with the sample surface, time-domain THz nanospectroscopy is performed by scanning the relative time delay between the optical sampling pulse and the scattered THz near-field pulse. This procedure maps the oscillating electric-field waveform, capturing both the amplitude and phase of the local response in the time domain. A subsequent Fourier transform of this time-domain signal extracts the broadband spectral response intrinsically correlated to the apex of the AFM probe.

For near-field imaging, the sampling delay is fixed at the peak of the THz pulse waveform to maximize the signal-to-noise ratio. The sample stage is then raster-scanned beneath the probe under continuous THz illumination. As the tip traverses the same topographic features in both the forward and backward directions, two independent data sets are generated with nearly identical spatial features. To further enhance image fidelity and suppress random noise, we perform a cross-correlation analysis between the forward and backward scans—leveraging the simultaneously acquired topography—to produce an averaged, high-resolution near-field map \cite{haeuser}.

\setcounter{figure}{0}

\renewcommand{\thefigure}{S\arabic{figure}}

\section*{Acknowledgments} Work at Ames National Laboratory was supported by the U.S. Department of Energy (DOE), Basic Energy Sciences, Division of Materials Sciences \& Engineering, under Contract No. DE-AC02-07CH11358. 

\section*{Author Contributions} 
J.W. conceived and supervised the project.
S.H., R.H.J.K., R.K.C., and J.M.K., performed the THz nano-imaging measurements. 
T.K., S.H., R.H.J.K., and M.M., developed the near-field model.  
S.H., and R.K.C., analyzed the data and performed simulations with discussions from all authors.  
The paper is written by S.H., and J.W., with input from all authors

\section*{Competing Interests} 
The authors declare that they have no competing financial interests.

\section*{Data Availability Statement} 
The data that support the plots within this paper and other findings of this study
are available from the corresponding author upon reasonable request.

	\bibliography{BibTex}
		
	\pagebreak

\end{document}